%%%%%%%%%%% TeX file starts from here
\magnification=\magstep1
\hsize 32 pc
\vsize 42 pc
\baselineskip = 24 true pt
\vskip .5 true cm
\def\cl{\centerline}
\cl {\bf A Quantum Mechanical Approach To A System of Self-Gravitating}
\cl {\bf Particles And The Problem Of Gravitational Collapse}
\vskip .5 true cm
\cl {\bf D.N. Tripathy and Subodha Mishra}
\vskip .2 true cm
\cl {Institute of Physics, Sachivalaya Marg, Bhubaneswar-751005,
Orissa, India}
%\vfill
%\eject
\vskip .5 true cm
\noindent {\bf Abstract}
\vskip .3 true cm
By making an intuitive choice for the single-particle density of a
system of N self-gravitating particles, without any source for
the radiation of energy, we have been able to calculate the binding
energy of the system by treating these particles as fermions. Our
expression for the ground state energy of the system shows a
dependence of $N^{7/3}$ on the particle number, which is in agreement
with the results obtained by other workers. We also arrive at a
compact expression for the radius of a star following which
we correctly reproduce the nucleon number to be found in a
typical star. Using this value, we obtain the well-known result for
the limiting value of the mass, $M$, of a neutron star $(M \simeq 3.12
M_{\odot}, M_{\odot}$ being the solar mass) beyond which the black
hole formation should take place. Generalizing the present
calculation to the case of white dwarfs,we have been able to
obtain the so called Chandrasekhar limit for the mass, $M_{Ch}$,
$(M_{Ch}\simeq 1.44 M_{\odot})$ below which the stars are expected to go over
to the white dwarf state. We reproduce this by introducing  a
radius, equivalent to Schwarzschild radius, at the 
interface of the neutron stars and white dwarfs. This is 
justified by considering the fact that it gives rise to the correct value
for the degree of ionization $\mu_e (\mu_e\approx 2)$ for heavy nuclei.
\vskip .4 true cm
Subject headings:Self-gravitating particles,Neutron star,Blackhole,

White dwarf, Chandrasekhar limit

\vfill
\eject
\vskip .4 true cm
\noindent {\bf 1. Introduction}

It was first pointed out by Chandrasekhar$^1$ and then,independently,
by Landau$^2$,long back that a degenerate system
composed of a large number of self gravitating particles will
necessarily undergo gravitational collapse if the particle number
exceeds certain critical value. This happens after the stars finish
up thier nuclear fuel. Soon after this, Chandrasekhar$^3$
made the momentous discovery regarding the life history of certain
stars, according to which the stars with masses $M$ less than $\approx
1.4 M_{\odot}, \ M_{\odot}$ being the solar mass,
evolve in the same way as the sun after the nuclear power in
their cores gets exhausted. When this happens, they contract to white
dwarfs. In a white dwarf star, the assembly of the free electrons
within the star, which usually forms a degenerate Fermi gas exerts
sufficient outward pressure to counteract the inward gravitational pull.
A star like our sun is said to lie on the main sequence of the 
Hertzsprung-Russel (HR) diagram$^4$, since it has still the source at
its core for the generation of energy. In the distant
future, it is also supposed
to evolve to become a red giant and then finally to a white dwarf.
Coming to the case of stars having masses less than about three times
the solar mass, they may condense even more as they collapse such
that their density becomes comparable to that of the nucleons inside
the atomic nuclei. At this stage, the electrons and protons react by
inverse $\beta$-decay and form neutrons. This is how the neutron
stars are formed. These are the compact objects having a dominance of
neutrons in their interiors. As such, in them, the outward pressure
arises from the degenerate neutrons. Lastly, one comes accorss the
most interesting case of stars that are having masses more than three
times the solar mass. In such cases, the collapse is complete and
they lead to the formation of the so called black holes. As the name
implies, the black holes trap light and material particles falling on
them and also prevent these from getting out of them. This is due to
the fact that gravity is very strong inside the black holes.
Mathematically, when the radius of a neutron star becomes less than a
certain limit called the Schwarzschild radius $R_s = ({2GM\over
c^2})$, $M$ being the mass of the star, $G$ Universial Gravitational
Constant and c, the velocity of light, then only it can become a
black hole. Since, it is the gravitational attraction among the
particles in a gravitating system which makes it to collapse,
this amounts to an
enormous increase in density and the temperature at its central
region. For a star becoming a black hole $(M\geq 3 M_{\odot})$, the
whole star enters the horizon and ends up as a singularity at the
centre. That is, the centre of a black hole is considered to be a
mathematical singularity where the matter is supposed to have infinite density.

It had been long since pointed out by Fisher and Ruelle$^5$ that to
have a rigorous treatment of an
infinite system of $N$ interacting particles using statistical mechanics, it is necessary that the
relevant forces must be of a saturating character. In that case, the total
energy of the finite system which is to be an extensive quantity
ought to possess a lower bound. That is, it is to be proportional to the number of
particle in the system. If, on the other hand, the forces are not of
saturating character, then the binding energy per particle increases
indefinitely with the number of particles $N$, so that it obviously
becomes impossible to define the usual thermodynamic variables for
such infinite systems. In a pioneering work Fisher and Reulle$^5$ have given the general criteria
to describe the saturation property for systems governed by not too singular
forces. An example of a non-saturating force is the well known
gravitational force which happens to be so, because of its long range
nature and attractive character. Taking all the facts into
consideration, Levy-Leblond$^6$ has succeeded in deriving both an
upper and a lower bound for the ground state energy of a
nonrelativistic quantum mechanical system of $N$ particles interacting
through gravitational forces. By treating these particles as bosons
he has shown that the binding energy per particle goes as $N^2$, whereas
for a system of fermions, it varies as $N^{4/3}$. However,by extending this
approach to a system consisting of $N$ negative light fermions with
mass $m$ and $N$ positive heavy fermions with mass $m_p$, $(m_p \ >> \ m)$ and treating
the entire system semirelativistically, the author$^6$ has found 
that the binding energy of the system also increases faster than $N$.
It is further shown by him that above a critical number of particles, $N_r$, the Hamiltonian is no
longer bounded from below and the system faces an unescapable
collapse. As an illustration of this calculation to the case of white
dwarf stars, he finds that the pressure of the degenerate electron
gas cannot balance the gravitational pull if the total number of
particles in a star is greater than the number $N_r = A({2\hbar
c\over G m^2})^{3/2}$, where A is some numerical coefficient which is to be
adjusted taking into consideration of the physics of the problem. The limiting value of
the mass of the star $M_r$ ($M_r = N_r m_n$,$m_n$ being the neutron mass) is being identified with the
socalled Chandrasekhar limit. Unfortunately, this very approach of
Levy-Leblond$^6$ cannot
be generalized to the case when  heavy particles,such as
neutrons, alone form a degenerate Fermi gas. Because, in that case, this
would need a full relativistic treatment of the gravitational
interaction. In a latter work, Ruffini and Bonazzola$^7$, without using
the equation of state approach, could succeed in doing a full relativistic calculation
of the binding energy of a system of N self-gravitating particles
each of mass $m_p$, $m_p$ being very heavy, following the self consistent
field method based on the general theory of relativity. By this they
were able to obtain a critical value for the particle number beyond which 
instability was found to set in within the system. However, for the number
of particles of such a high order of magnitude, it was shown by them that the Newtonian
treatment of such a system led to an utter failure.

We, in the present work, have tried to calculate the binding energy of
a self-gravitating system of particles by treating them as fermions.
As far as the evaluation of the total kinetic energy of the
system is concerned, it is done within the Thomas-Fermi approximation (TF)$^8$. The
potential energy of the system is being evaluated within the socalled
Hartree approximation. The form of the single-particle density for
the system used by us in the present calculation is such that it has
a singularity at the origin. Unlike the earlier calculations, the
method used here is a nonrelativistic quantum mechanical derivation
based on Newtonian mechanics. The most interesting result of the
present theory is that it gives rise to a compact expression for the
radius of a star, following which we are able to obtain a limiting value for
the critical mass of a neutron star in a natural way beyond which the
black hole formation takes place. A further generalization of
the present work to the case of white dwarfs enables us to derive
the socalled Chandrasekhar limit. In sec.2 of
this paper we have presented the mathematical formulation of our
theory. Sec.3 is devoted to the various situations
which lead to the formation of neutron stars, and black holes and the
derivation of the well-known Chandrasekhar limit for the white
dwarfs. In sec.4 a brief discussion of the results of the present theory is given.
\vskip .3 true cm
\noindent {\bf 2. Mathematical Formulation of the Theory}

The Hamiltonian for the system of $N$ gravitating particles each of
mass $m$ interacting through a sum of pair-wise gravitational
interactions is written as
$$ H = \sum^N_{i=1} - {\hbar^2\nabla^2_i\over 2m} + {1\over 2}
\sum^N_{i=1} \sum^N_{j=1} v (\mid\vec X_i -\vec X_j\mid),\eqno{(1)}$$
where $v (\mid\vec X_i-\vec X_j\mid) = -g^2/\mid\vec X_i-\vec X_j\mid,$
with $g^2 = Gm^2$, $G$ being Newton's Universal gravitational constant.
Using this, the ground state energy of
the system at zero temperature is given as$^8$
$$ E_0 = <H> = <KE> + <PE>,\eqno{(2a)}$$
Assuming the particles to be fermions,the total kinetic energy of the
system $<KE>$ has been evaluated within the Thomas-Fermi
approximation,whose expression is given as
$$<KE> = ({3\hbar^2\over 10m}) (3\pi^2)^{2/3}\int d\vec X [ \rho
(\vec X)]^{5/3},\eqno{(2b)}$$
and the total potential energy $<PE>$ is written as
$$ <PE> = -({g^2\over 2})\int d\vec X d\vec X' {1\over \mid\vec X-\vec
X'\mid} \rho (\vec X)\rho (\vec X')\eqno{(2c)}$$
In order to evaluate the above integrals, we assume that all the
particles within the system (which are identical in nature)are described by some kind of
distribution. The trial single-particle density $\rho (\vec X)$ we
choose is of the form:
$$\rho (\vec X) = A [exp(-x^{\alpha})]/ x^{3\alpha},\eqno{(3a)}$$
where
$x = (r/\lambda), r = \mid \vec X \mid$ and A is the
normalization constant, such that
$$ \int\rho (\vec X) d\vec X = N \eqno{(3b)}$$
The index $'\alpha'$ has been adjusted inorder to bring the
expression for the binding energy to have the correct dependance with
the particle number.
Besides, the convergence of the integrals is also to be 
satisfied. As one can notice from above, $\rho (\vec X)$ is singular
at $r = 0$. The existence of a singularity in the single-particle
density at the origin of the co-ordinate system need not be
unphysical. In case of a black hole it simply means that the centre
of a black hole is a mathematical singularity where matter has
infinite density. As far as the universe is concerned, a
singularity in the particle density at origin is thought to be
related to the so called Big Bang theory, which is being assumed to
be the most important current theory for the origin of the Universe.
There have been a few most important advances in this direction by Hawking and
Penrose$^4$  who have shown that any model of the Universe which has
the observed characterstics of approximate homogenity and isotropy
must start from a singularity. Even, Einstein's General Theory of
Relativity (GTR) which when applied to cosmology accounts for such an initial
singularity of the Universe.

  Evaluations of the integrals in Eq.(2) have been made taking a
set of values for $\alpha$ like $\alpha =
4,3,2,1$ and $\alpha = {1\over 4}$,${1\over 3}$ and ${1\over 2}$. It
is to be noted that the value $\alpha = (1/2)$ proves to be the
most appropriate choice for the single particle density of the
system. This can be seen from the results we shall be discussing later.
With help of the above choice for $\rho (r)$, 
the expression for the total energy ,$E_0(\lambda)$,of the system of $N$
self-gravitating particles is obtained as
$$E(\lambda) = \bigg({12\over 25\pi}\bigg) \bigg({\hbar^2\over m}\bigg)
\bigg({3\pi N\over 16}\bigg)^{5/3} {1\over\lambda^2} - \bigg({g^2 N^2\over
16}\bigg) {1\over \lambda}\eqno{(4)}$$
Minimizing this with respect to $\lambda$, it is found that the
minimum occurs at
$$\lambda = \lambda_0 \simeq ({\hbar^2\over mg^2}) \times (2.023764)/N^{1/3}\eqno{(5)}$$
Evaluating Eq.(4) at $\lambda = \lambda_0$,the total binding
energy of the system is found as 
$$E_0 \simeq - ( 0.015442 ) N^{7/3} ({mg^4\over\hbar^2})\eqno{(6)}$$
Considering the case of the two-particle system (N=2), from Eq.(6), we find
$$ E_0 = - (0.077823)({mg^4\over \hbar^2})$$
This is seen to be quite high compared to the actual binding energy
of the two-body system whose value is (-0.25) $({m g^4\over\hbar
^2})$. Comparing the two results, one should not consider Eq.(6) to be
a drawback of the present theory,because it is supposed to be very
accurate for very large $N$. Looking at Eq.(6), we find that $E_0$
varies as $N^{7/3}$ where $N$ is the particle number.
Such a dependence of the binding energy for the system on
$N$ was also found by Levy-Leblond$^6$ by assuming the
particles to be fermions and looking at the distribution of N-points
on a cubic lattice. By this, he was able to obtain both an upper and a
lower bound for the binding energy of the system which, for large N,
were given as$$ - (0.5) N^{7/3} ({m g^4\over \hbar^2}) \leq E_0\leq -
(0.001055) N^{7/3} ({m g^4\over \hbar^2})\eqno{(7)}$$
Anyway,comparing our result, as shown in Eq.(6), with Eq.(7), we find that it does
not violate the inequalities established by Levy-Leblond$^6$.

%\vfill
%\eject 
\vskip .3 true cm
\noindent {\bf 3. Formation of Compact Objects}
\vskip .2 true cm
\item {3.1.} {\bf Neutron Stars and Black Holes}
\vskip .1 true cm
 Before we go to make an estimate of the critical mass of a neutron
star beyond which black hole formation should take place,we have to
first know about the radius of a star. It must be noted that the size
of any compact object (either an atom or a star) is not well defined
in quantum theory. The justification regarding the identification of
the radius $R_0$ of a star with $2\lambda_0$ follows from the
consideration of the socalled tunneling effects used in quantum mechanics.
Classically,it is known that a particle has a turning point where the
potential energy becomes equal to the total energy$^9$.Since the
kinetic energy and therefore the velocity are equal to zero at such a
point,the classical particle is expected to be turned around or
reflected by the potential barrier. For example,considering the case
of an electron in the hydrogen
atom ground state such classical turning point occures where the
potential $V (r) = -e^2 /r  = E_{total} = -e^2/2a_0$ ;that is at $r=2
a_0$. Quantum mechanically,the probability distribution $r^2\rho (r)$
has a non-zero value for $r>2 a_0$ ; that is, the electron has access
to the region $r>2 a_0$ which is forbidden by classical theory. Such
penetration or tunneling into or through the potential energy
barriers is typical of quantum theory results. If the electron had a
value of $r >2a_0$,then its kinetic energy would have to be negative
to satisfy the condition $E_{total} = T + V$ ,with $V > E_{total}$.
Since negative kinetic energy is physically absurd, $r =2a_0$ is to
be identified as the classical radius. Using the above idea, from the
present theory one can easily see that at $\lambda =
2\lambda_0$, the potential energy of the system becomes equal to the
the total energy,there by proving that the radius of the star $R_0 = 2\lambda_0$.

In order that a neutron star, after it finishes up all its
nuclear fuel at the centre, would form a black hole, one must have
$$R_0 \leq R_s, \eqno{(8)}$$
where
$$R_s = ({2GM\over c^2}),\eqno{(9)}$$
is the Schwarzschild radius$^4$ of the corresponding black hole.
Following Eqs.(5) and (8), one finds that the number of nucleons N in the
star satisfies the inequality
$$ N \geq  1.696758 N_1,\eqno{(10 a)}$$
where
$$ N_1 = ({\hbar c\over G m_n^2})^{3/2},\eqno{(10 b)}$$
$m_n$ being the mass of a neutron. The equality sign in
Eq.(10 a) refers to the critical value, denoted by $N_{C_1}$, for the
number of particles in a neutron star beyond which black hole
formation takes place. A numerical estimation of $N_{C_1}$ gives
$$N_{C_1}\simeq 1.70 N_1\approx 3.73 \times 10^{57}\eqno{(11)}$$
From Eq.(10 b) it follows that
$$ N_1^{2/3} = ({\hbar c\over Gm_n^2}) = \bigg ({Plank mass\over nucleon
mass}\bigg )^2\eqno{(12)}$$
Looking at the result given in Eq.(11), one finds that this is in fantastic agreement with the
well known result for the number of nucleons in a typical star, 
as estimated earlier$^{10}$. Using this, one also finds that
$$M_{C_1} = N_{C_1} m_n\simeq 3.122134 M_{\odot},\eqno{(13)}$$
where $M_{\odot} = 2 \times 10^{33} g,$ is the mass of the sun. Thus, we find that for
neutron stars more massive than $\approx 3 M_{\odot}$, the collapse
is complete and these are the stars which lead to the black holes$^4$.
Now, corresponding to $N = N_{C_1}$, we calculate the radius of a neutron star, which gives

$$R_0 = 2\lambda_0 \leq {3.39352 ({\hbar \over m_nc}) ({\hbar c\over Gm_n^2})^{1/2}} \eqno{(14)}$$
This is the same result as found earlier by Shapiro and
Teukolsky$^{11}$ (ST). A numerical estimate of Eq.(14) gives $9.25\times
10^5cm$ compared to the value of $3\times 10^5 cm$ quoted by ST

\vskip .3 true cm
\noindent {3.2.} { \bf White Dwarfs and The Derivation of the Chandrasekhar limit}

In view of the result shown in Eq.(13), it is apparent that if the mass of a star is less
than $\approx 3 M_{\odot}$, but not too low, it must remain as a
neutron star.
At this stage, one is likely to ask, is there any lower bound on the
mass of a neutron star ? In order to answer this question, we imagine
of a radius, denoted by $R_s'$, equivalent to the Schwarzschild radius, upto which the
neutron star is likely to exist. Above this 
$R_s'$, one no longer talkes of a neutron star. Rather, one has to
speak of a white dwarf, provided the mass of the star is less than
the Chandrasekher limit at the time when its nuclear fuel gets exhausted.
Mathematically, we write down
the expression for the $R_s'$ as
$$R_s' = {2 GM\over <\vec v^2>}\eqno{(15)}$$
As one can see from above, $R_s'$ has been written in a fashion similar to the
Schwarzschild radius except for the fact that the $c^2$ factor in the
Schwarzschild radius has been replaced by the average of the
velocity square $<\vec v^2>$. The quantity  $<\vec v^2>$ is to be
here understood as the escape velocity of a particle from a neutron
star. Quantitatively, we choose $<\vec v^2>$ as
$$ <\vec v^2> = c^2 \bigg ({m_e\over m_n}\bigg )^{\eta},\eqno{(16)}$$
where the value of the exponent $\eta$ in the above equation is to be adjusted inorder
to reproduce the value 2 for the degree of ionization for heavy
nuclei. In doing this,the socalled Chandrasekhar limit$^3$ for the
mass of a white dwarf ($M\approx 1.44 M_{\odot}$) is obtained in a
natural way. In order to show this, we now consider the following
inequality,
$$R_s < R_0 < R_s' = \bigg [ {2 GM\over c^2}\bigg ] \bigg ({m_n\over
m_e}\bigg )^{\eta},\eqno{(17)}$$
Analysing Eq.(17), for the case $R_0 < R_s'$, we obtain
$$N = N_{C_2} \geq 1.696757 \bigg({\hbar c\over Gm_n^2}\bigg)^{3/2}
\bigg[\bigg({m_e\over m_n}\bigg)^{3/4}\bigg]^{\eta} \eqno{(18)}$$
Following this, we write
$$M_{C_2} = m_nN_{C_2} \geq 3.126 M_{\odot} \bigg[3.5613 \times
10^{-3}\bigg]^{\eta} \eqno{(19)}$$
Using the above equation, we now go on varying $\eta$. For each value
of $\eta$, we try to calculate the degree of ionization $\mu_e$ using
the relation$^{12}$
$$\mu_e^2 = 5.83 ({M_{\odot}\over M_{C_2}}) \eqno{(20)}$$
It can be easily seen that only when $\eta = 0.137271$, $\mu_e$
becomes 2.01. For heavy nuclei,it has been known that $\mu_e$,which
is being interpreted as the degree of ionization has a value close to
2. Now,corresponding to the above $\eta$,we find that 
$$M_{C_2} \simeq 1.44 M_{\odot},\eqno {(21)}$$
the well known Chandrasekhar limit$^2$.
A further justification regarding our above choice of
$R_s'$ is given in sec.4. Thus, the mass of a neutron
star happens to be such that $M_{Ch} \leq M^{NS} \leq 3.12 M_{\odot}$. For
a star having masses $M < M_{Ch}$, the formation of white dwarfs
should take place after such a star finishes up all its nuclear energy. 

In order to calculate the radius of a white dwarf star, one has to
consider the fact that in these stars, the outward pressure is due to
the degenerate electrons rather than due to the neutrons as is the
case with the neutron stars. Therefore, in white dwarfs, it is this
outward electron pressure which is counterbalanced by the inward
gravitational pull arising out of the neutrons. While generalizing
the present calculation to white dwarfs, we ignore the effect of
the gravitational forces between the electrons and electrons and
between electrons and neutrons, as these are negligibly small. This
is justified considering the fact that the neutron mass is very
high compared to the electron mass. Thus the mass $`m'$ that appears in
the kinetic energy term in  Eq.(1) should now represent the electron mass
$m_e$ and the symbol $g^2$ that appears in the interparticle
potential term should, as before, be given as $g^2 = G m^2_n$, $m_n$
being the mass of a neutron. With these modifications, the expression
for $R^{WD}_0$ is obtained as
$$R^{WD}_0 \simeq \bigg ({\hbar^2\over G m_e m^2_n}\bigg ) 4.047528/
N^{1/3}\eqno{(22)}$$
$R^{WD}_0$ should be such that its value has to
be greater than $R_s'$. It can be easily verified that for masses
$M\leq 1.44 M_{\odot}$, $R_0^{WD} \ > \ R_s'$.
For $ M = 1.0 M_{\odot}$, we have calculated the radius of the white
dwarf using  Eq.(22). This gives $R^{WD}_0\approx 2.49 \times 10^9$ cm.
which is in close agreement with the value estimated by others$^{11}$.
For this mass, $R_s'\approx 0.832\times 10^6cm$; thus showing that
$R_0^{WD} > R_s'$. Using the above value of $R^{WD}_0$, we have estimated the mass density
inside a white dwarf of mass $M = 1.0 M_{\odot}$. This gives
$\rho^{WD}\approx 3.1 \times 10^4 g/cm^3$, which is again
of the right order of magnitude as reported by others$^{13}$.
Using the above value of $\rho^{WD}$, the density of
particles within a white dwarf star is found
to be $\approx 1.80 \times 10^{28} \ cm^{-3}$. It is because of such
a high value for the particle density, the effects of the pauli exclusion principle
becomes important in such stars and hence, the matter in such a state is
considered to be quantum mechanically degenerate.

 Since $R_0^{WD}$ is also supposed to be larger than $R_s$, this
gives rise to the fact that 
$$N \geq 1.696757 \bigg({\hbar c\over Gm_n^2}\bigg)
\bigg({m_n\over m_e}\bigg)^{3/4} \eqno{(23)}$$
Following this, one obtains
$$R_0 \geq 0.5184 ({\hbar \over m_ec}) ({\hbar c\over Gm_n^2})^{1/2} \eqno{(24)}$$
This is the well known relation as obtained before$^{11}$. The
expression in the right hand side of the above equation when
evaluated gives $\approx 2.6\times 10^8cm$. For $M=1.0 M_{\odot}$,
the estimated value of $R_0^{WD}$ actually satisfies the above
ineqality. Now, consider the case of a neutron star of mass $M = 1.5 M_{\odot}$. Following
Eq.(5), its
radius becomes $R_0 = 1.18 \times 10^6$ cm. Using this, the
matter density inside such a star is found to be $\rho^{NS} \approx
4.3 \times 10^{14} g/cm^3$. This being of the same order as the mass
density within an atomic nucleous, one is justified to call them as
neutron stars. In the black hole state $(M \simeq 3.2 M_{\odot})$,
the radius of the corresponding neutron star becomes $R_0 = 9.18
\times 10^5 cm$. Its Schwarzschild radius $R_s$ is found to be 
$3({M\over M_{\odot}}) km \simeq 9.6 \times 10^5 cm$. Thus, one finds
that for a black hole, $R_0 < R_s$. This is what is expected to happen
for neutron stars having masses $M \geq 3.12 M_{\odot}$.

Coming back to the case of a star in the white dwarf stage with a
mass $ M \simeq 1.0 M_{\odot}$, we have estimated the mean temperature
throughout the body of such a star by requiring that the thermal
kinetic energy of the star be equal to its gravitational potential
energy. Using the present theory, we have calculated the total binding
energy of a white dwarf star of mass $M = 1 M_{\odot}$, using the expression
$$\mid E_0\mid = 0.015442 N^{7/3} {G^2 m_e m^4_n\over\hbar^2},\eqno{(25)}$$
This gives
$$\mid E_0\mid \approx 0.67 \times 10^{49} \ ergs$$
Comparing our result with those estimated earlier$^{14}$,
we find that the agreement is extremely good. Leaving aside the
details of the composition, as far as the star like the sun is
concerned, which is at present a star on the main sequence, it can be considered
to resemble with a white dwarf after its nuclear fuel gets exhusted.
Using virial theorem, which tells
that the sum of the potential energy and twice the kinetic energy of
a self-gravitating system is zero$^4$, we obtain
$$\mid E_{pot}\mid \simeq 1.34 \times 10^{49} ergs,\eqno{(26)}$$
Following  Eq.(26), we now calculate the value of the potential energy
per gramme, and then equate it with the mean thermal kinetic energy,
${1\over 2} v^2$, per gramme of a particle (Hydrogen atom) inside the
white dwarf. This gives the mean thermal velocity $v$ of the particle
as $v \simeq 1.6 \times 10^3 km/sec$, and hence, the corresponding
mean temperature becomes $\sim 5.5 \times 10^7$K . The
central temperature of a white dwarf has to be much more than the
above value. As far as the sun is concerned, since its binding is
found to be less than that of a white dwarf of the same mass$^{14}$,
it is expected that the mean temperature of the sun has to be less
than that of a white dwarf. The same is true for the central
temperature which, for the sun, has a value $\sim 2 \times 10^7 K$.
\vskip .3 true cm
\noindent {\bf 4. Discussion}

As seen before, in order that a star can go to white dwarf state
after its nuclear fuel at the core gets exhausted, it must have a mass
less than $\approx 1.44 M_{\odot}$, the well known Chandrasekhar
limit. To arrive at this result, we have introduced a radius
$R'_s$, equivalent to the Schwarzschild radius $R_s$, such that $R_0 <
R_s'$, where we have defined $R_s' = {2 GM\over < \vec
v^2>},$ having $<\vec v^2> = c^2 ({m_e\over m_n})^{\eta},$ which is
being interpreted as the escape velocity of a particle from the
surface of the neutron star. In order that the above inequality is to be
satisfied, one must have $M < M_{Ch} \approx (1.44) M_{\odot}$,
which corresponds to an $\eta = 0.137271$.
For such an $\eta$, we reproduce a value for the
escape velocity $v (v \approx 0.62 c)$ of a particle from the surface
of the neutron star which is found to be of the right order of magnitude$^{15}$.
This also gives the correct result for the degree of
ionization, $(\mu_e\approx 2)$ 

To conclude, we, in this work, have succeeded in obtaining a
nonrelativistic quantum
mechanical derivation for the ground state binding energy
of a system of self-gravitating particles by making a suitable choice
for the single particle density. This has enabled us to
arrive at a compact expression for the radius of astronomical
objects like stars. Using the present theory, we have been able to estimate the
critical mass of a neutron star beyond which black hole formation
takes place. The present derivation of the Chandrasekhar limit for the
white dwarf formation is based on introducing a radius equivalent to
the Schwarzschild radius at the region of interface between the white
dwarfs and neutron stars. From all these successes we
feel that our choice for the single particle density of a system of
self-gravitating particle is correct. Further investigation on
various other properties of the system are under progress.
\vfill
\eject
\noindent {\bf References}
%\item {}  Bransden,B.H. and Joachain,C.J. 1983, Physics of atoms and molecules (New York:Longman)
%\item {}  Chandrasekhar,S. 1931, Monthly Notices Roy.Astron. Sco.91,456
%\item {}  Chandrasekhar,S. 1935, Monthly Notices Roy.Astron. Sco.95,207
%\item {}  Contopoulos,G. and Kotsakis,D. 1987, Cosmology (Heidelberg:
% Springer-Verlag)
%\item {}  Fisher, M.E. and Ruelle,D. 1966 ,J. Math. Phys. 7, 260
%\item {}  Flower,P. 1990,Understanding the Universe(St.paul:West Publ.Comp.)
%\item {}  Harrison,E.R. 1981, Cosmology (Cambidge: Cambridge University Press)
%\item {}  Landau, L. 1932, Phys. Z. USSR 1, 285
%\item {}  Levy-Leblond,J.M., 1969 ,J. Math. Phys. 10, 806
%\item {}  Lipunov,V.M. 1992 ,Astrophysics of Neutron Stars(Berlin:Springer-Verlag)
%\item {}  Long,C.E. 1943,Discovering the universe(New York: Harper and Row)
%\item {}  Ruffini,R. and Bonazzola,S.1969,Phys. Rev. 187 , 1767.
%\item {}  Sciama,D.W. 1971, Modern Cosmology,
%(Cambridge: Cambridge University Press)
%\item {}  Shapiro,S.L. and Teukolsky, S.A. 1983, Black Holes,
%White Dwarfs and Neutron Stars (New York: John Wiley and Sons,Inc.)

\item {[1]} S. Chandrasekhar, {\it Monthly Notices Roy.Astron. Sco}. {\bf 91},456 (1931).
\item {[2]} L. Landau, {\it Phys. Z. USSR}  {\bf 1}, 285 (1932).
\item {[3]} S. Chandrasekhar, {\it Monthly Notices Roy.Astron. Sco}. {\bf 95},207 (1935).
\item {[4]} G. Contopoulos and D. Kotsakis, {\it Cosmology},

(Heidelberg,  Springer-Verlag, 1987),pp. 10,125,127,93,100,56.
\item {[5]} M.E.Fisher and D.Ruelle ,{\it J. Math. Phys.}, {\bf 7}, 260 (1966).
\item {[6]} J.M. Levy-Leblond, {\it J. Math. Phys.}, {\bf 10}, 806 (1969).
\item {[7]} R.Ruffini and S. Bonazzola, {\it Phys. Rev.}, {\bf 187} , 1767 (1969).
\item {[8]} B.H.Bransden and  C.J.Joachain, {\it Physics of atoms and

molecules},(New York,Longman,1983),pp.353.
\item {[9]} M.Karplus and R.N.Porter, {\it Atoms and

Molecules},(Reading,Massachusetts: W.A.Benjamin,Inc.1970),pp.115-116.
\item {[10]} E.R.Harrison, {\it Cosmology}, (Cambidge: Cambridge University Press,1981),pp.334.
\item {[11]}S.L.Shapiro and S.A.Teukolsky, {\it Black Holes,
White Dwarfs and Neutron Stars}, (New York, John Wiley and Sons,Inc.1983),pp.65.
\item {[12]}V.M.Lipunov, {\it Astrophysics of Neutron Stars},
(Berlin:Springer-Verlag,1992),pp.2.
\item {[13]}C.E.Long, {\it Discovering the universe},(New York, Harper and Row,1943),pp.334.
\item {[14]}D.W.Sciama, {\it Modern Cosmology},(Cambridge,Cambridge University

 Press,1971),pp.3-4.
\item {[15]}P.Flower, {\it Understanding the Universe},(St.paul,West Publ.Comp.1990),pp.576.
\vfill
\eject
\end